\documentstyle[aps,prb,multicol,epsf,graphicx]{revtex}

\newcommand \bi{\bibitem}

\newcommand \be{\begin{equation}}
\newcommand \ee{\end{equation}}
\newcommand \beq{\begin{eqnarray}}
\newcommand \eeq{\end{eqnarray}}

\begin{document}

\title{Strong Soret effect in one dimension}
\author{A.Garriga$^{2}$, J.Kurchan$^{1}$ and F.Ritort$^{2}$}      
\address{
$^{1}$ P.M.M.H. Ecole Sup\'erieure de Physique et Chimie Industrielles\\
10, rue Vaquelin,75231 Paris CEDEX 05, France\\
$^{2}$Department de F\'{\i}sica Fonamental,
Facultat de F\'{\i}sica, Universitat de Barcelona\\
Diagonal 647, 08028 Barcelona, Spain\\
}

\maketitle
\begin{abstract}
We consider a one-dimensional gas of two kinds of particles with
different masses interacting through short range interactions.  The
system exhibits an extreme form of the Soret effect: when the ends of
the system are in contact with thermal baths of different
temperatures, there is complete separation of the species. We show how
this separation can be well described in the Boltzmann approximation
and discuss the origin of this odd behavior.
\end{abstract}
\section{INTRODUCTION}

When a binary fluid is subjected to a temperature gradient, the
densities $\rho_+$ and $\rho_-$ of the species become space-dependent
in such a way that the relative concentration $\rho_+(x)/\rho_-(x)$
changes along the direction of heat flow.  This is the Ludwig-Soret,
or Soret effect \cite{Soret}.  That this effect happens does not in
itself need an explanation: once the temperature difference takes the
system away from equilibrium and breaks translational symmetry, there
is no reason why the ratio of concentrations should stay constant.

However, the Soret effect has practical applications in the separation of
species \cite{Martin}, and this requires a quantitative
 prediction of its magnitude in each case.
There have been several analytic approaches to do this for gases
 \cite{Chapman}, condensed phases \cite{deGroot},  grains in
 suspension \cite{Ruckenstein}, porous media
 \cite{dG}, etc. (For a review, see \cite{Grew}).

From a purely theoretical point of view
the Soret effect is interesting because it cannot in general  be discussed
in terms of local equilibrium, in which a space-dependent local temperature
 fully accounts for the concentration ratio --- as for example the
 local pressure accounts for the variation of the density
 in a gas column under gravity.
To see this explicitly, we consider here a model belonging to
 a family of systems for which such a local equilibrium 
approximation yields strictly zero effect, contrary to observation.
They consist of two species of
different masses $m_+$ and $m_-$, interacting through a pair potential 
$V(r_i-r_j)$ which does not depend on the particle type 
(the masses of particles  $i$ and $j$). 
A short calculation shows that the partition function $Z$  of such a system
is of the form:
\be
\frac{1}{V} \ln Z(\rho_+,\rho_-,T) =- \rho_+ \ln \rho_+
-\rho_- \ln \rho_- + 
\frac{d}{2}  \rho_+ \ln(m_+ T) + \frac{d}{2}  \rho_- \ln(m_- T) -
\beta F(\rho, T)
\label{cosa}
\ee
where $F$ depends on $\rho_+ $, $\rho_-$ only through their sum
$\rho \equiv \rho_+ + \rho_-$,
and $d$ is the dimension.
Assume now that under heat flow 
 we can divide space in small cells $\delta^d(x)$ with densities
 $\rho_+(x)$ and
$\rho_-(x)$, 
 within which the system is in  equilibrium at temperature $T(x)$.
The total partition function will then be a product 
expressions  (\ref{cosa}) for each cell. If we now maximize it
  with respect to
$\rho_+(x)$ and $\rho_-(x)$ (with fixed  number of particles)
we get:
 \begin{eqnarray}
\alpha_+ &=& 
-\frac{1}{T(x)} \frac{\partial F}{\partial \rho}(\rho, T(x)) - \ln(\rho_+)
+  \frac{d}{2}  \ln(m_+) \nonumber \\
\alpha_- &=& 
-\frac{1}{T(x)} \frac{\partial F}{\partial \rho}(\rho, T(x)) - \ln(\rho_-)
+  \frac{d}{2}  \ln(m_-)
\end{eqnarray}
with $\alpha_\pm$ space-independent Lagrange multipliers.
Subtracting these equations we get $\rho_+/\rho_- = constant$, i.e. no
Soret  effect.

In this paper we study a very simple one-dimensional model which
belongs to the class described above, and has a most extreme form of
Soret effect: there is total phase separation even for arbitrarily
small temperature differences.  The two pure phases are separated by
an interface in which species are mixed, its width is a function of
the parameters.  Although this full separation is not realistic, it is
interesting to exhibit a model which, for large sizes, is driven far
from equilibrium by an arbitrarily small difference of temperature.

We also study in this model the closely related question of the laws for
  heat transfer. In  regions in which
phases are pure this transfer has an anomalous spatial and temperature
dependence,  just as found
 in oscillator chains \cite{chains}. Within the interface we show that 
the behavior becomes the usual Fourier law, as recently found by 
Dhar \cite{Dhar} in a model similar to the present one,
but in which particles are forced to stay mixed.

\section{THE MODEL}

We shall consider a one dimensional system of length $ L $ 
consisting of $ N $ point particles of two species: $ N_+ $ 
heavy and  $ N_- $ light particles of masses $ m_+ $ and $ m_- $, respectively.

Particles interact through an
 infinitely narrow potential of height $V$, so that
 when two particles meet they collide if their  center of mass energy 
\be
E_c = 
\frac{1}{2} \frac{m_1m_2}{m_1+m_2} (v_1-v_2)^2 < V
\ee
 and ignore each other otherwise.
The collision between particles conserves energy and momentum:
\be
v_1 \longrightarrow v_1'=v_1+\frac{2m_2}{m_1+m_2}(v_2-v_1)
\ee
\be
v_2 \longrightarrow v_2'=v_2-\frac{2m_1}{m_1+m_2}(v_2-v_1)
\ee
These collision rules are reversible and satisfy detailed balance.
Between collisions the particles evolve freely across the
system. Particles colliding against the walls $ x=0 $ and $x=L$
rebounce with a velocity with random distribution corresponding to
thermalisation at temperatures $T_c$ and $T_h$, respectively:
\be
P_{\alpha}(v) dv=\frac{m}{\kappa T_{\alpha}}v \exp
\left(-\frac{mv^2}{2\kappa T_{\alpha}} \right)dv~~~\alpha=c,h
\label{wall}
\ee
This guarantees 
that if $T_c=T_h$ the system equilibrates to the Maxwell distribution.

In fact, only the potential between particles of different
 species is relevant, since  particles of equal mass only exchange their 
velocities in a collision, which then just amounts to exchanging their
 labels --- a one dimensional peculiarity. 
As a consequence of this, 
we can consider this model as having a  potential $V(r_i-r_j) $ which
 is independent of particle type (and hence  belongs to the family
 (\ref{cosa})), and at the same time consider that particles of the
 same type are 
 transparent to one another, and hence our model is a particular case
 of the one introduced by  Widom and Rowlinson\cite{Row}.

\section{PHASE SEPARATION}

We use a simple molecular dynamics in the simulation, calculating the
minimum  collision time and letting  the system evolve freely
between collisions. We perform simulations for different number of
densities, and different values of the length ranging from $ L=100 $
to $ L=2000 $, which allows us to consider finite-size effects.  

We have checked that in the equilibrium case (the temperature in both
sides of the system are identical) the two species are homogeneously
distributed in space and their velocity distribution is Maxwellian. 
We
also have checked that for the case of one type of particles with two
different temperatures, the average energy is the one corresponding
to an homogeneous temperature $ \sqrt{T_cT_h} $.
 This is because if we
consider only one type of particles, in the elastic case the particles
simply cross the system without any interaction with other particles
and it is easy to verify that the distribution of velocities is just
composed of two Maxwellian distributions  at different temperatures
for particles flying in each direction.

\begin{figure}[tbp]
\begin{center}
\includegraphics*[width=14cm,height=8cm]{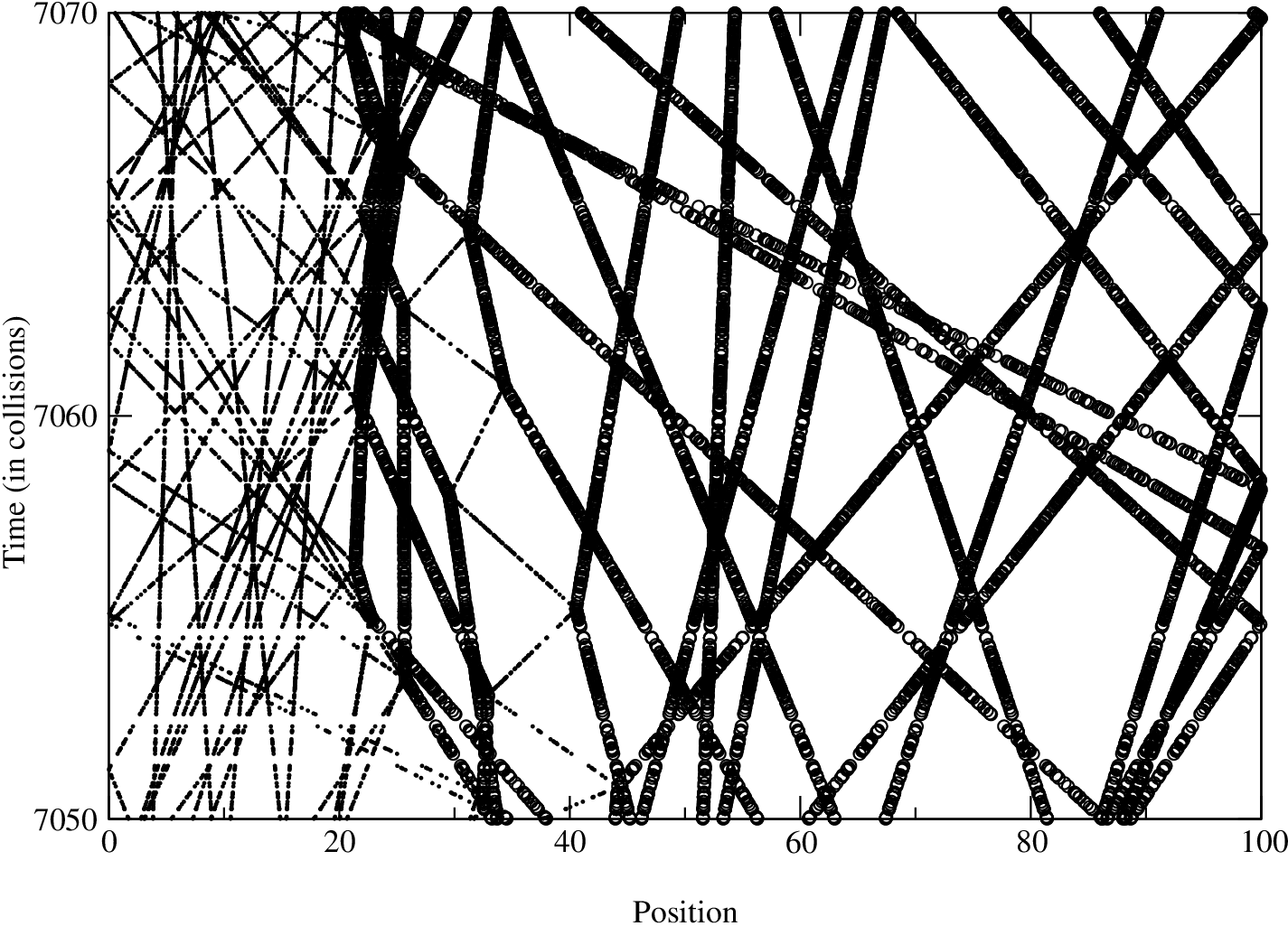}
\vskip 0.1in
\caption{Plot of the time evolution of the positions
 of the particles. The large  circles correspond to a
 particles with  $ m_+=1 $ while the small ones to a 
 $ m_-=0.2 $.
 The walls are a $ T_c(x=0)=1 $ and  $ T_h(x=100)=10 $.
 The interaction potential is $ V=10 $. There are 20 particles of each type. 
\label{fig1}}
\end{center}
\end{figure}

In Fig.\ref{fig1} we show the evolution of the particles after a long
thermalisation time, starting from a configuration with random
positions ($ L=100 $, $ V/T_c=10 $, $T=T_h/T_c=10$, $ m_-/m_+=0.2 $).
We can see that the space is divided in two regions with only a single
species, separated by an interface in which both species mix.
Away from the interface, because all particles have the same mass,
the system can be considered as non-interacting.
Hence we have the following picture: the heavy particles 
collide with the hot wall 
 at $x=L$. They fly unperturbed (modulo relabelings)  throughout the
 the heavy-particle phase, until they meet the first light particles
 around the interface. After one or more crossings and collisions with 
light particles, they return
to the heavy-particle phase, where they are free again.
The same can be said of the light particles on the left.

A simple computation helps clarify what happens in the interface. 
Consider a system with a light and a heavy particle.
Estimating their velocities according to  the temperatures of the
walls,
we have:

\begin{itemize}
\item If the light particle is on the left and the heavy one on the right:
\be
\left< v_-^2 \right> \approx \frac{T_c}{m_-}~~~
\left< v_+^2 \right> \approx \frac{T_h}{m_+}~,
\ee
which gives a typical center of mass energy:
\be
2E_{CM}^{(1)}=\frac{(T_c+T_h)(m_++m_-)-(\sqrt{T_cm_-}+\sqrt{T_hm_+})^2}{(m_++m_-)}~.
\ee 
\item If the heavy particle is on the left and the light one on the right:
\be
\left< v_-^2 \right> \approx \frac{T_h}{m_-}~~~
\left< v_+^2 \right> \approx \frac{T_c}{m_+}~,
\ee
giving the typical precollision center of mass  energy:
\be
2E_{CM}^{(2)}=\frac{(T_c+T_h)(m_++m_-)-(\sqrt{T_hm_-}+\sqrt{T_cm_+})^2}{(m_++m_-)}
\ee
\end{itemize}
In fact is easy to show that:
\be
A=\frac{2E_{CM}^{(1)}}{2E_{CM}^{(2)}}~ <~ 1~~~ {\rm{if}}~~~ m_+~ >~
m_-;~ 
T_c~<~T_h~,
\label{asy} 
\ee
Hence, we see that the probabilities of collision are not symmetric:
while this does not prove that phase separation exists, it gives us
the sign of the effect.
Note that there is a similarity with asymmetric exclusion
 processes \cite{AEM}, although unlike that case here  detailed 
balance holds in the bulk. 

\section{INTERFACE}

\begin{figure}[tbp]
\begin{center}
\includegraphics*[width=14cm,height=7cm]{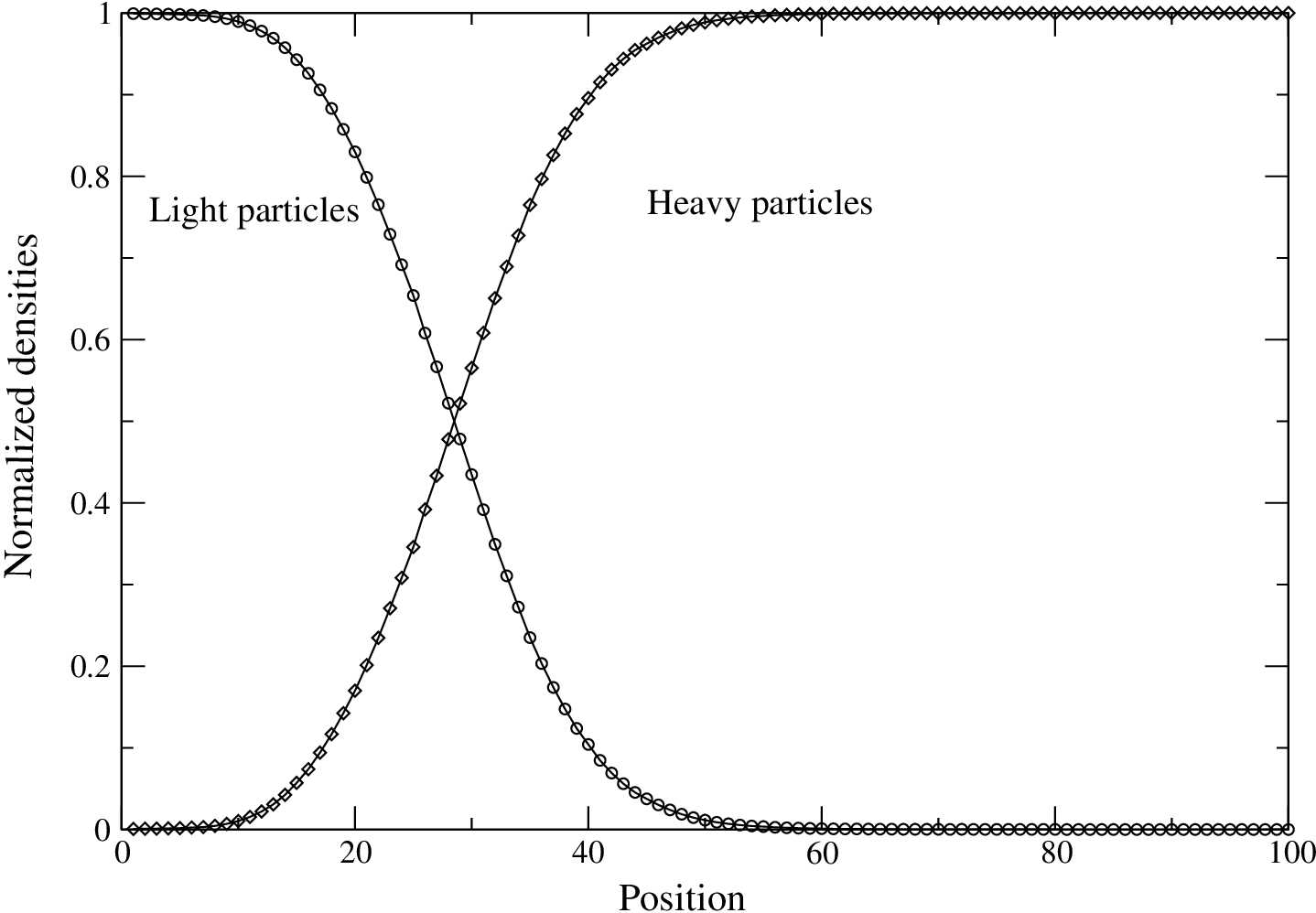}
\vskip 0.1in
\caption{Plot of the densities $\rho_+$ and $\rho_-$.  The parameters
of the model are the same as in Fig.\ref{fig1}.  Data obtained with
100 bins taking measures every 10 collisions during a time
corresponding to 500.000 collisions per particle. The diamonds
correspond to the heavy particles while the circles correspond to the
light ones.
\label{fig2}}
\end{center}
\end{figure}

In Fig.\ref{fig2} we show the time-averaged particle densities
$\rho_+(x)$ and $\rho_-(x)$: \be \rho_\pm(x) = 1/t \lim_{t \rightarrow
\infty} \int_0^t \; dt' \sum_{i \in \pm} \delta(x-x_i(t')) \ee for a
system with the same parameters as Fig.\ref{fig1}. We see, as before,
that the two types of particles are completely separated, but the
interface seems much thicker than in Fig.\ref{fig1}.  The reason for
this discrepancy is simple: the position of the interface fluctuates
with time.  Upon time-averaging we are actually measuring the
instantaneous interface width convoluted with the dispersion in its
position.
\vspace{0.75cm}

\begin{figure}[tbp]
\begin{center}
\includegraphics*[width=14cm,height=8.5cm]{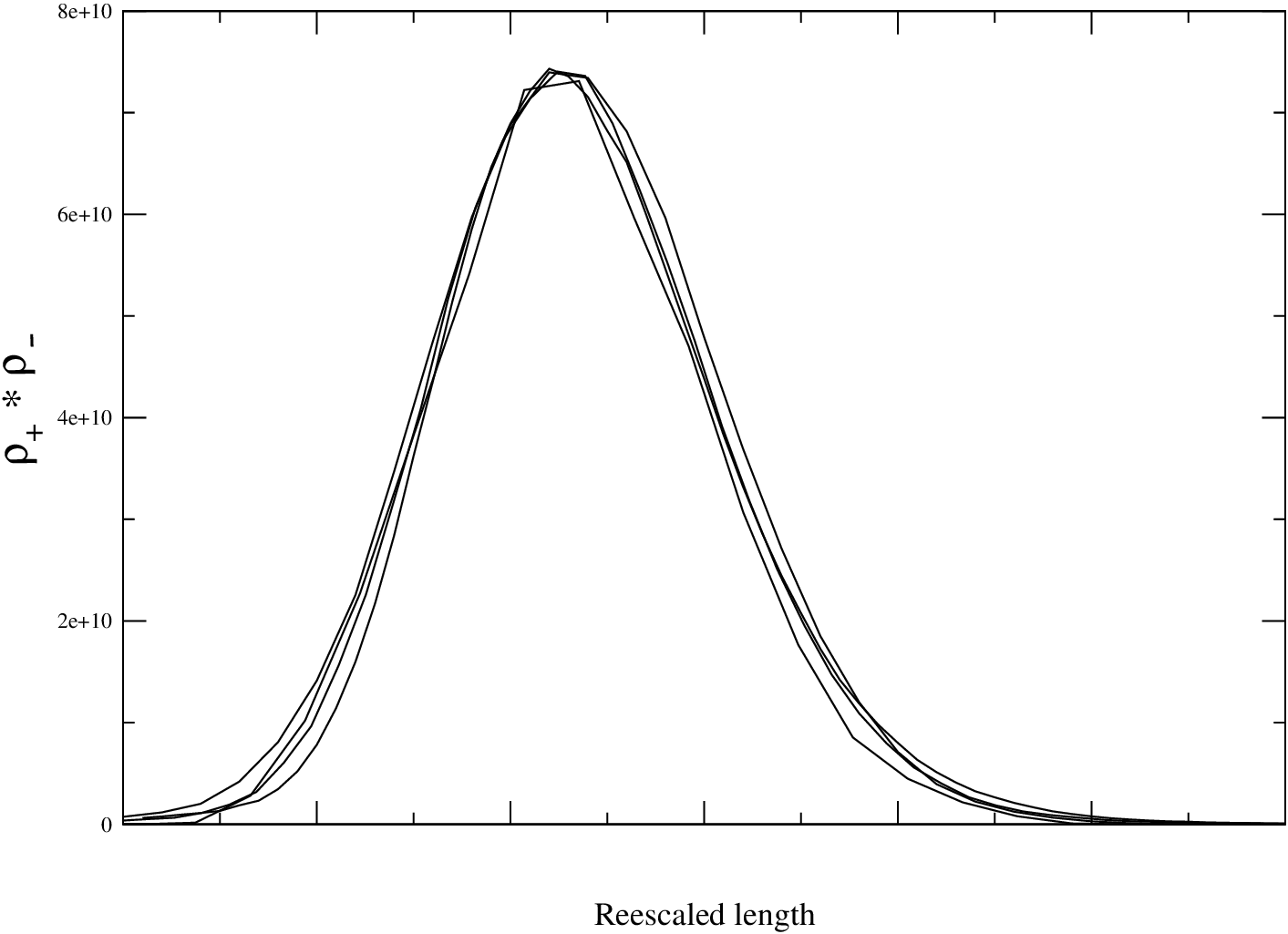}
\vskip 0.1in
\caption{Plot of the product $\rho_+(x)\rho_-(x)$.  For different
sizes $L$, centering and rescaling the interface with $ \sqrt{L} $.
The curves correspond to $ L=100,200,400,800 $.
\label{fig3}}
\end{center}
\end{figure}

The fluctuations in the position of the interface $\Delta L$ scale
  with the system size $L$. In Fig.\ref{fig3} we check the assumption
  that, just like in an ordinary equilibrium system $\Delta L \propto
  \sqrt{L}$. In order to do this, we plot the product of densities
  $\rho_+(x)\rho_-(x)$, a quantity that is large only in the
  interface, for several system sizes and we verify that, indeed, the
  function scales with $\sqrt{L}$.
 In \ref{fig3} we have considered again the same parameters as in
  Figs.\ref{fig1} and \ref{fig2}, and systems lengths from $ L=100
  $ to $ L=800$. Rescaling the averaged interface with $\sqrt{L}$, the
curves  collapse.

\begin{figure}[tbp]
\begin{center}
\includegraphics*[width=14cm,height=8cm]{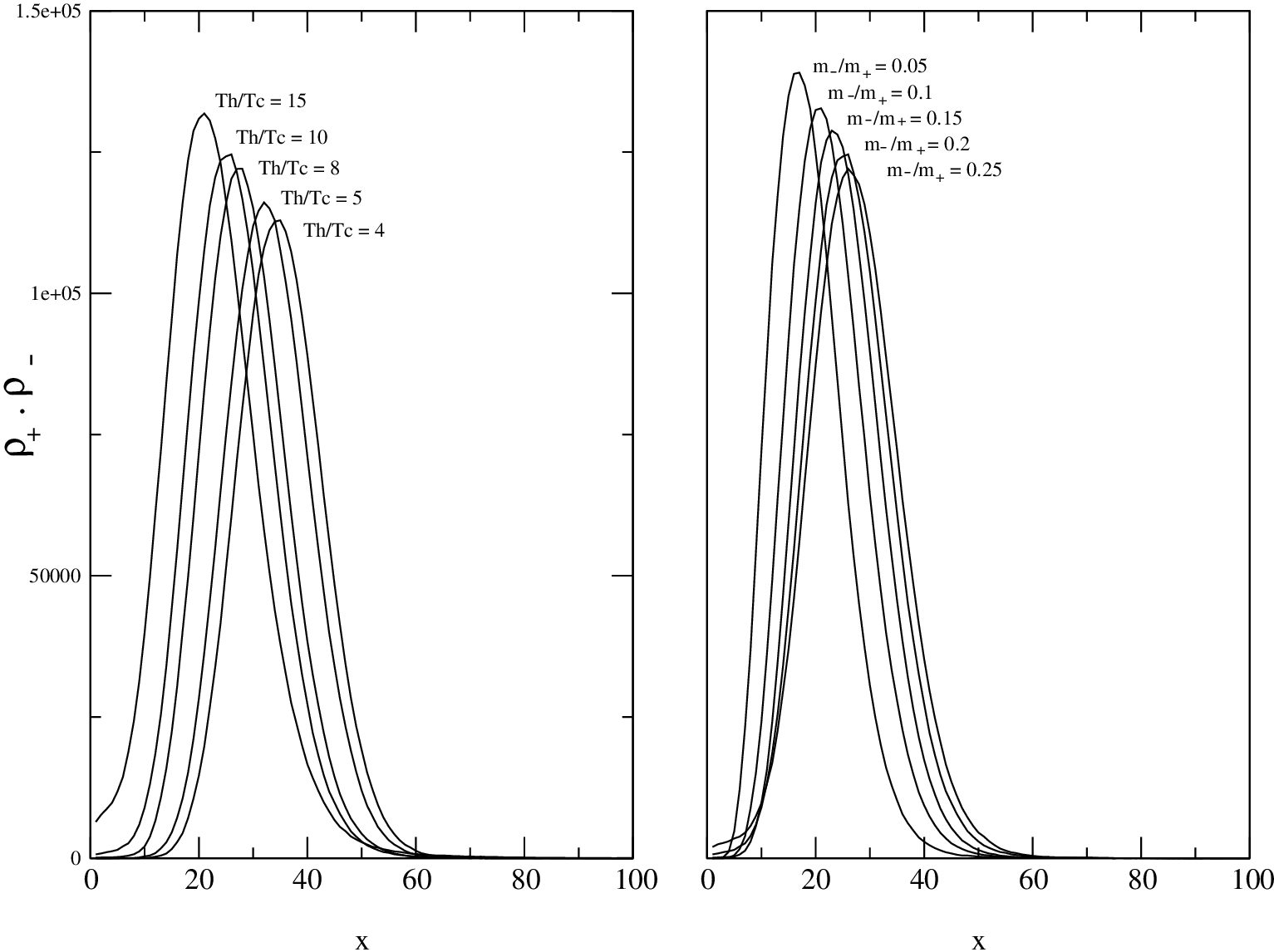}
\vskip 0.1in
\caption{Product $\rho_+(x)\rho_-(x)$ for different values of the
parameters.  Left: $ T_h/T_c = 4,5,8,10,15 $ and $ m_-/m_+ = 0.2 $.
Right: $ m_-/m_+ = 0.05, 0.1, 0.15, 0.2, 0.25 $ and $ T_h/T_c =10$.
\label{fig4}}
\end{center}
\end{figure}

 In Figures \ref{fig4} we plot $\rho_+\rho_-$ for a system of length
 100 and $ V=10 $.  On the left figure we vary the temperature of the
 hot wall keeping the mass ratio constant, while on the right figure
 we have changed the mass ratio while keeping the temperatures
 constant.  These plots show us the shift in terms of the parameters
 of the average location of the wall.
     
\begin{figure}[tbp]
\begin{center}
\includegraphics*[width=14cm,height=8.5cm]{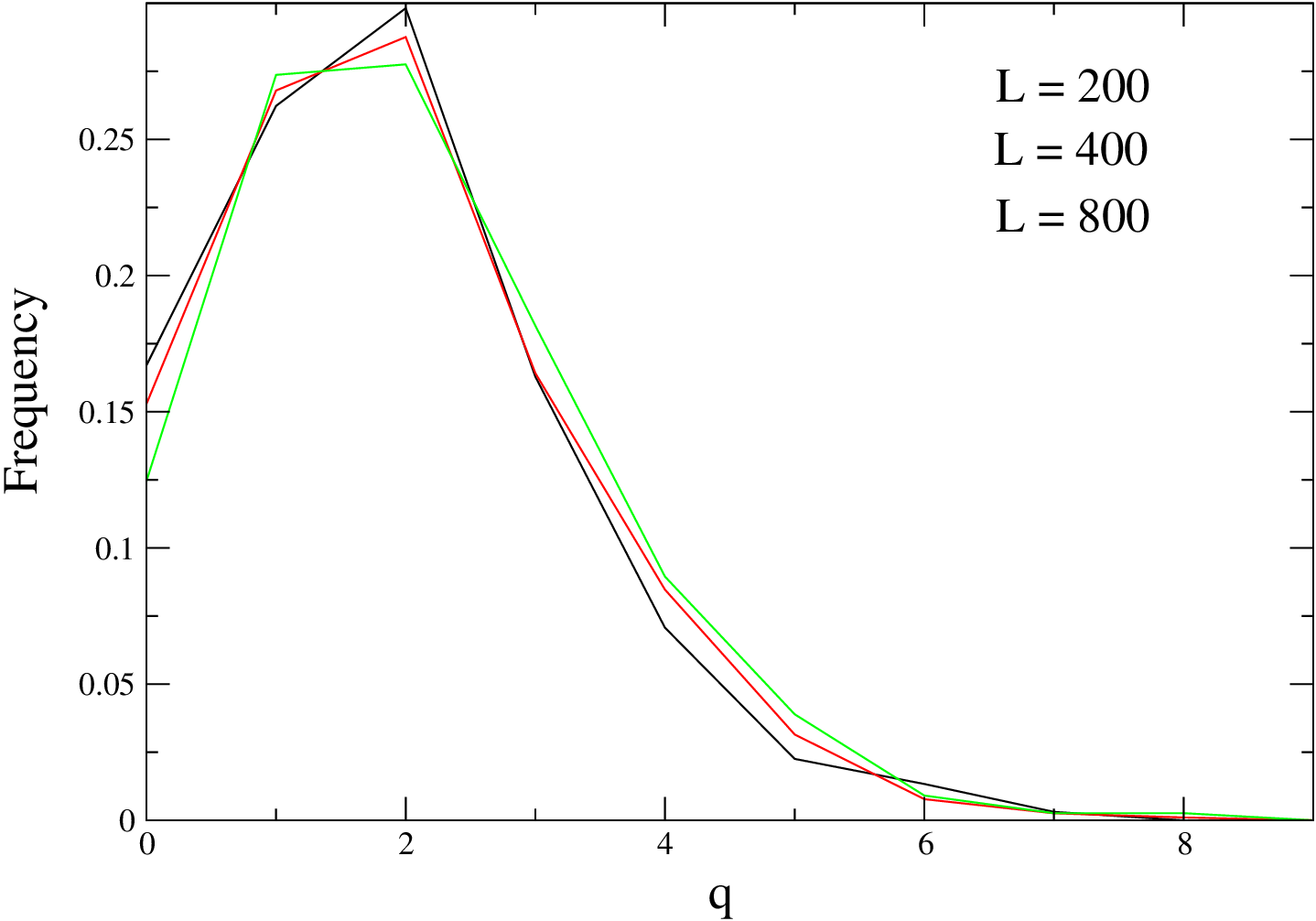}
\vskip 0.1in
\caption{Check of the size independence of the averaged frequency
of  $ q $. The parameters are
 $ V/T_c = 5 $, $ T_h/T_c = 10 $,  $ m_-/m_+ = 0.2 $.
\label{fig5a}}
\end{center}
\end{figure}

 As we  seen above, the
 time-averaged width of the interface is dominated by the time
 fluctuations of the wall position, rather than by its instantaneous
 width. Hence, it is convenient to
 define a parameter that reflects the mixing of the particles at
 every time, thus giving a measure of the instantaneous width of the
 interface.  For each heavy particle we count the number $q_i$
 ($=0,1,2$) of its nearest neighbors that are light particles.  Our
 parameter is then: \be q \equiv \frac{\sum_i^N q_i -1}{2} \ee For a
 sharp interface $q=0$, and for a very mixed system $q$ is of order
 $N$.  In Fig.\ref{fig5a} we show the frequency distribution of $ q $
 averaged over many time-steps for various sizes of the system. We see
 that the averaged distribution of $q$ (which reflects the
 instantaneous width of the interface) is independent of the system
 size, as expected.  We can hence study how the width changes with the
 various parameters, in a system-size independent way.

 On (a) we can see how the sharpness of the interface increases as we
 decrease the mass ratio. The same effect occurs when we increase the
 interaction potential (b) or the temperature ratio (c).
The physical interpretation of these tendencies
is easy to understand. The system is
completely mixed when the two thermal walls are at the same
temperature. In fact, in this case the asymmetry (\ref{asy}) in the
probability of collisions disappears ($A=1$), and equilibrium is
reached. Obviously if the masses are equal or the interaction
potential is zero, the system is also completely mixed because there
are no collisions.

  We
 have found no evidence of a threshold in the values below which there
 is no separation for an infinite system. Indeed, the limit of `mixed
 particles' (e.g. at equal temperatures or equal masses) is only achieved
 when the interface length becomes larger than system size.

\begin{figure}[tbp]
\begin{center}
\includegraphics*[width=16cm,height=8cm]{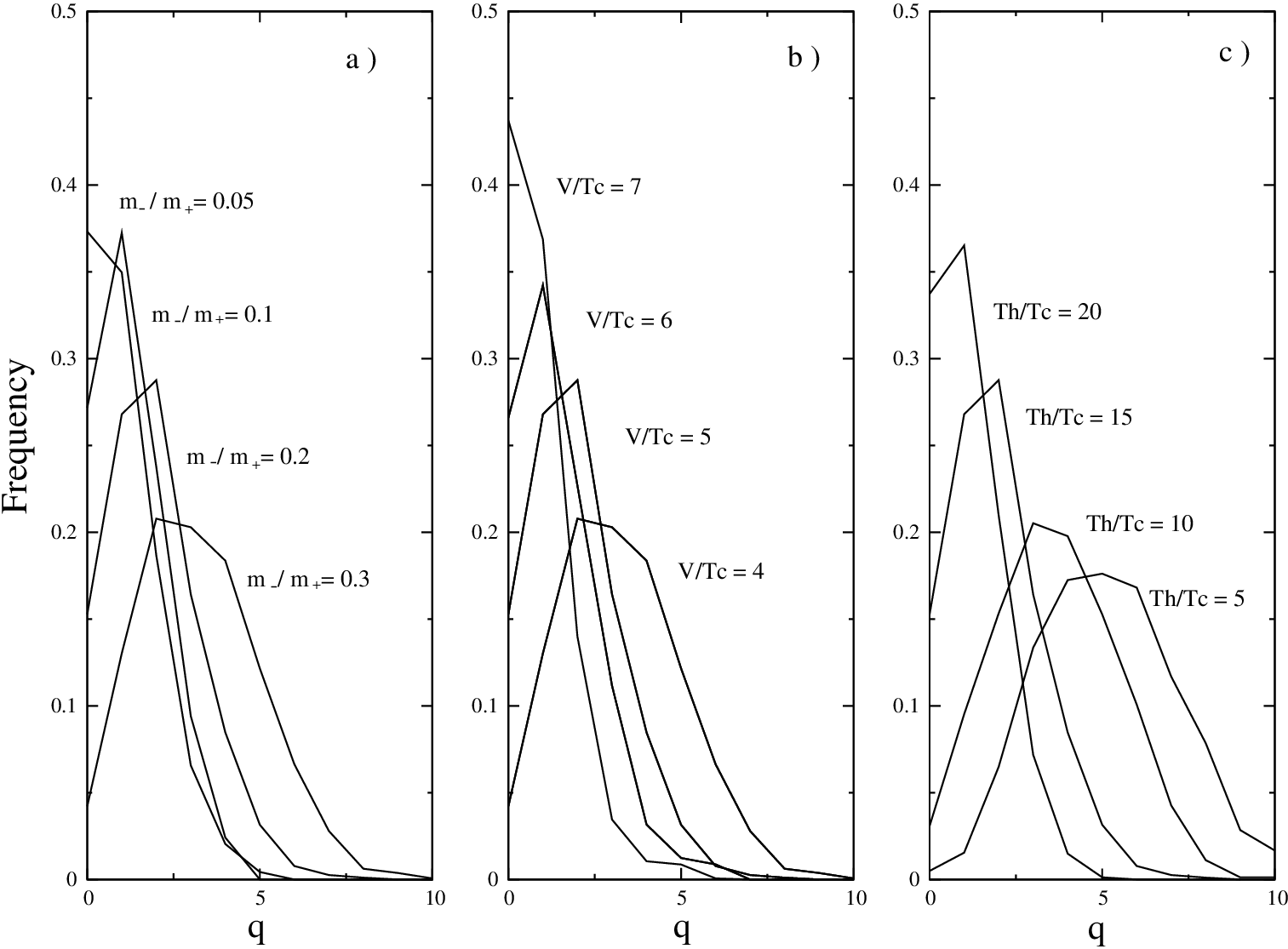}
\vskip 0.1in
\caption{Dependence of the distribution of  $ q $ on the parameters.
(a) $Th/Tc=10$, $m_-/m_+=0.2$ and $V/T_c=4,5,6,7$;
(b) $Th/Tc=10$, $V/T_c= 5$ and $m_-/m_+=0.05,0.1,0.2,0.3$
(c) $m_-/m_+=0.2$, $V/T_c= 5$ and $Th/Tc=5,10,15,20$
\label{fig5b}}
\end{center}
\end{figure}

\section{BOLTZMANN EQUATION}

Let us now write down a Boltzmann equation
 for the probability distribution function of
 the velocities of the particles, and check that it reproduces 
correctly the  phase-separation effect. 
 
We have to deal with two  coupled equations:
\be
\frac{\partial f(x,v,t)}{\partial t}=
-v\frac{\partial f}{\partial x} + 
\int |v-u|( f(x,v',t) g(x,u',t)- f(x,v,t) g(x,u,t))du
\label{boltz1}
\ee
\be
\frac{\partial g(x,v,t)}{\partial t}=-v\frac{\partial g}{\partial x} + 
\int |v-u|( f(x,u',t) g(x,v',t)- f(x,u,t) g(x,v,t))du
\label{boltz2}
\ee where $ f(x,v,t) $ and $ g(x,v,t) $ are the probability
distributions for the two types of particles. As usual, $ u' $ and $
v' $ are velocities the particles should have before the collision in
order that the velocities after collision are $ u $ and $ v $
respectively.  These equations must be solved with the adequate
boundary conditions (\ref{wall}). In this equations we have omitted
the interaction potential, in fact there would be a $\Theta$-function
which guarantees the collision. 

 The approximation that is made in equations (\ref{boltz1}) and
(\ref{boltz2}) is that the correlations between particles are
neglected: each particle collision is independent of the collisions
the particle has suffered before. More formally, this means that: \beq
f_2(x,x',v,v',t)= f(x,v,t) f(x',v',t)\\ g_2(x,x',v,v',t)= g(x,v,t)
g(x',v',t) \eeq Unfortunately, as far as we know, the system
(\ref{boltz1}) and (\ref{boltz2}) cannot be solved analytically.  We
solve it numerically using a stochastic process based on the Bird
algorithm \cite{BIRD} which has been proved \cite{WAG} to converge to
the solution of the Boltzmann equation. We have used this algorithm
for this system taking into account the delta-like potential.

\begin{figure}[tbp]
\begin{center}
\includegraphics*[width=12cm,height=7cm]{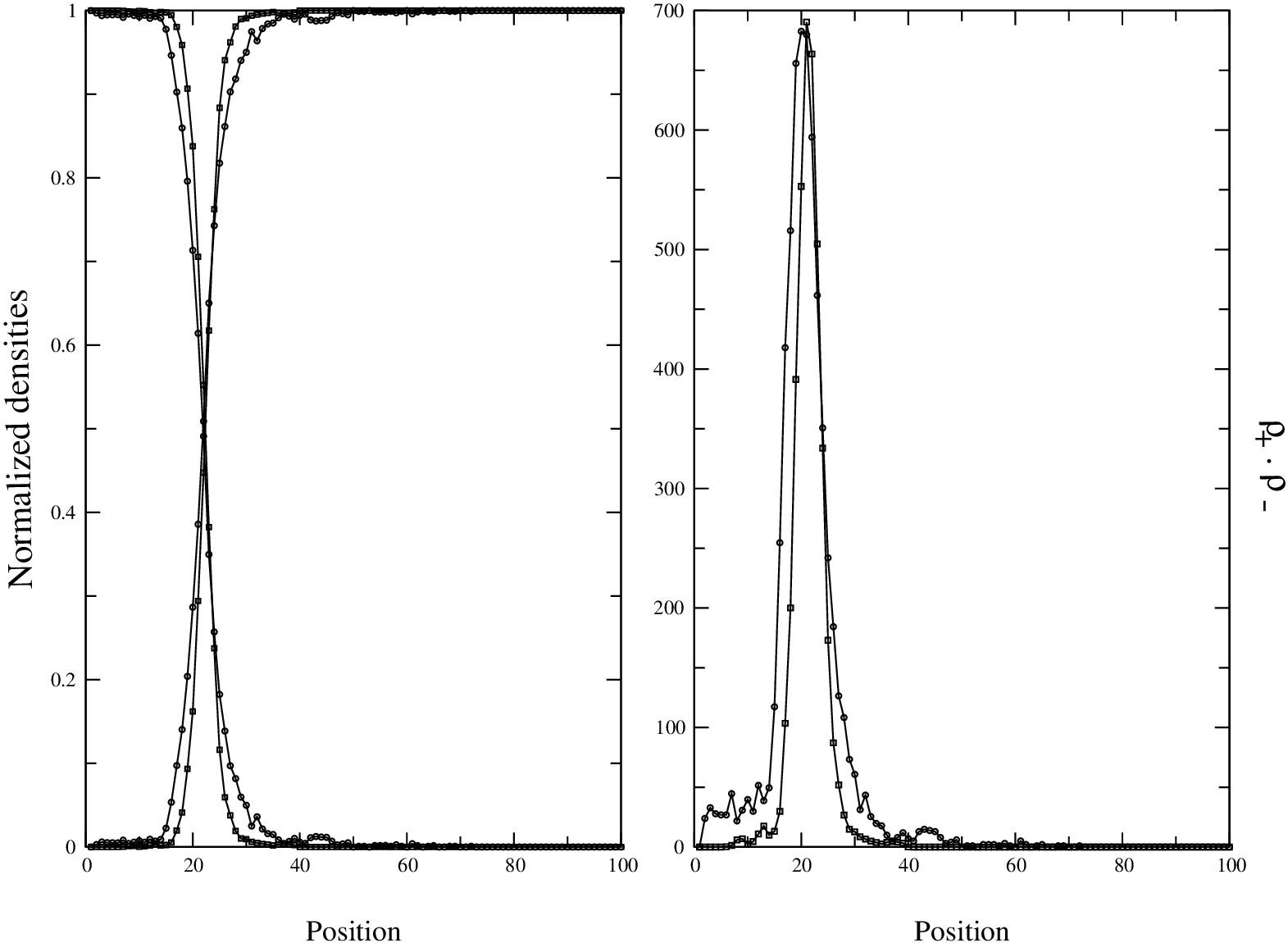}
\vskip 0.1in
\caption{ Left: Densities profiles (left) and $\rho_+(x)\rho_-(x)$
  (right)
for a
  system
 of length $100$ with $200$ particles of each type and
 $ V/T_c=4 $, $ m_-/m_+=0.2 $ and $ T_h/T_c=10 $. 
 The squares represent  molecular dynamics while the circles
  the results of the Boltzmann equation.  
\label{fig6}}
\end{center}
\end{figure}

 In Fig.\ref{fig6} we show the profile of the spatial density of the
two types of particles for a system consisting of 200 particles of
each type for a potential $ V/T_c=4 $, mass ratio $ m_+/m_-=0.2 $ and
temperature ratio $ T_h/T_c=10 $. We can see that the two normalized
profiles are almost identical, which confirms the validity of the
Boltzmann approximation. This good agreement holds for any value of the
parameters.

The fact that the Boltzmann equation is a good approximation, at least
away from the interface, is easy to understand.  As the two species
are separated, the most likely situation after a collision is that at
least one of the two colliding particles goes freely to its
corresponding wall, it loses all memory of the correlation with other
particles. Thus, the neglect of the correlations implicit in the
Boltzmann equation is a good approximation.

A special note should be made about the interface fluctuations.  The
 Boltzmann equation has a unique solution for every finite length $L$
 (it is first order). However, if we were to compute the stability
 matrix around the solution, we would find that the mode corresponding
 to translation of the wall becomes softer as $L \rightarrow
 \infty$. This in turn means that corrections to the Boltzmann
 approximation become more and more important in that limit, and they
 yield the wall's fluctuations.  (This question becomes more familiar
 if we bear in mind the analogy with a 2D Ising ferromagnet with the
 spins on right boundary set to $+1$ and on the left to $-1$. The
 Boltzmann equation is analogous to a mean-field solution, which
 places the interface in the middle, and neglects the fluctuating
 displacements of the domain wall.)

\section{HEAT TRANSPORT - RECOVERY OF FOURIER LAW FOR BROAD INTERFACE}
 
 In one dimensional systems, the obtention of Fourier's law is still
 an open question \cite{leb}. As mentioned above, this one dimensional
 model has the peculiarity that particles can be considered
 non-interacting everywhere except in the interface, where both
 species mix.  This reflects itself in an extremely pathological form
 of heat transfer: the only spatial gradient in kinetic energy of the
 particles occurs across the interface, since far from the
 interface  we have a  gas of noninteracting particles.
 It thus seems reasonable, if we wish to recover the usual laws
 for heat transfer, to place ourself in conditions such that the
 interface is broader than the sample itself and species are mixed.
               
 In order to study the temperature profile in the interface we have
chosen very small temperature differences and species with very
similar masses.  In Fig.8 we show the energy profile for a system of
length $ L=100 $ with $100$ particles of each type, with masses $
m_+=1 $ and $ m_-=0.95 $, respectively.  We plot the energy profile
for $ m_+ $ for different values of the temperature ratio, $ T_h/T_c $=
1.05, 1.10, 1.15, 1.20. We can clearly see that the energy
distribution changes linearly along the system, and that it is
proportional to the thermal gradient. In fact, the straight lines
fitted have all the same slope, indicating that the conductivity in
the interface is the same for all cases.

\begin{figure}[tbp]
\begin{center}
\includegraphics*[width=14cm,height=8cm]{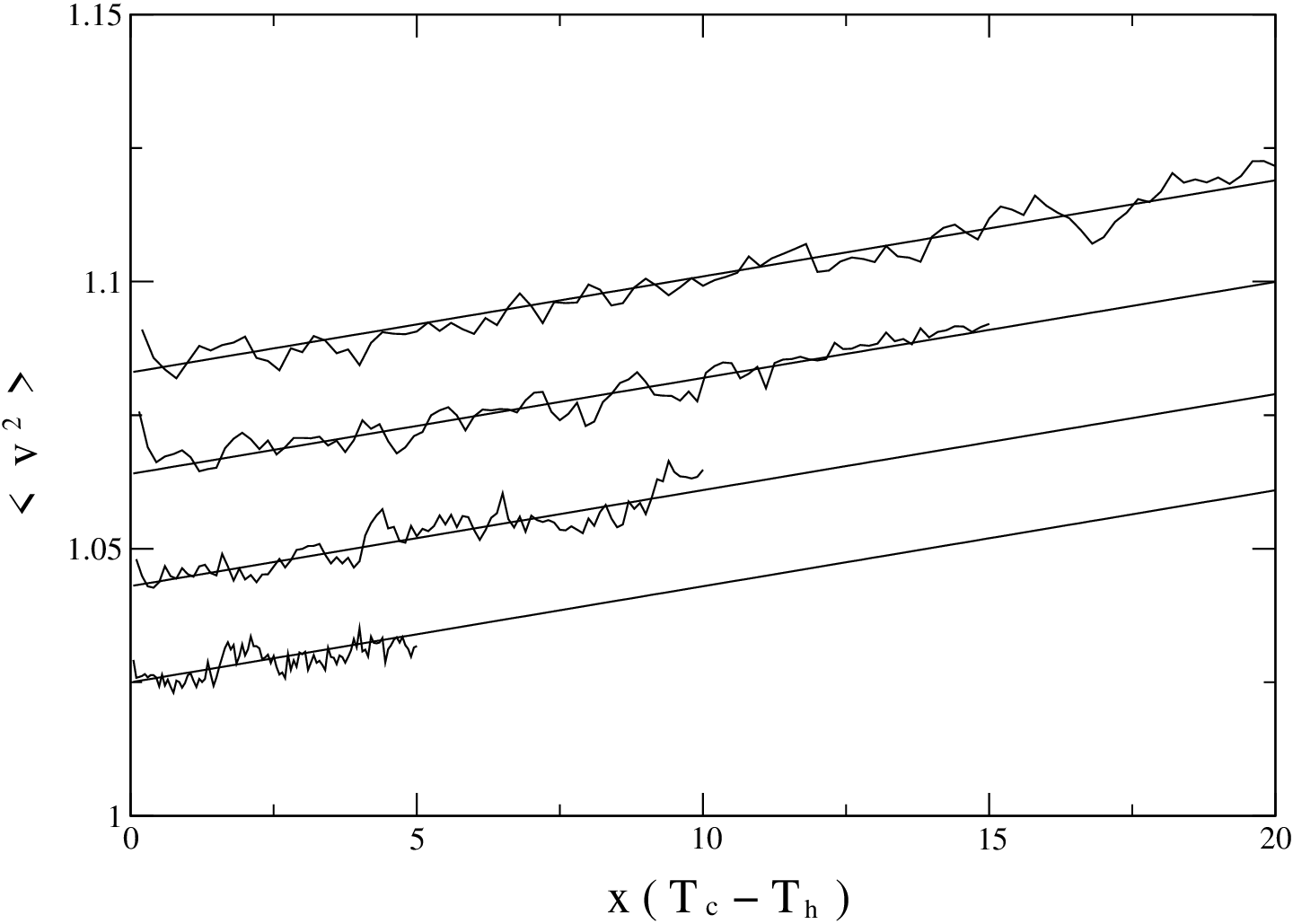}
\vskip 0.1in
\caption{Plot of the energy profile in function of the temperature
ratio along the system for the particles with mass $ m_+=1 $ when the
other particles have a mass of $ m_-=0.95 $. The length of the system
is $ L=100 $. The temperature ratios plotted are, from top to bottom ,
$ T_h/T_c $= 1.20, 1.15, 1.10, 1.05. All the fitted lines have the
same slope.
\label{fig8}}
\end{center}
\end{figure}  

Even in this small gradient case we can see that the mass distribution
is not uniform. In fact, the massive particles are located closer to
the hottest wall while the lightest ones are closer to the coldest
wall. This fact can be seen in Fig.9, where we have plotted the
normalized  particle densities for the same case as in Fig.8 for $ T_h/T_c
= 1.30 $.  Note that the mass profile is also linear in small thermal gradients.

\begin{figure}[tbp]
\begin{center}
\includegraphics*[width=14cm,height=8cm]{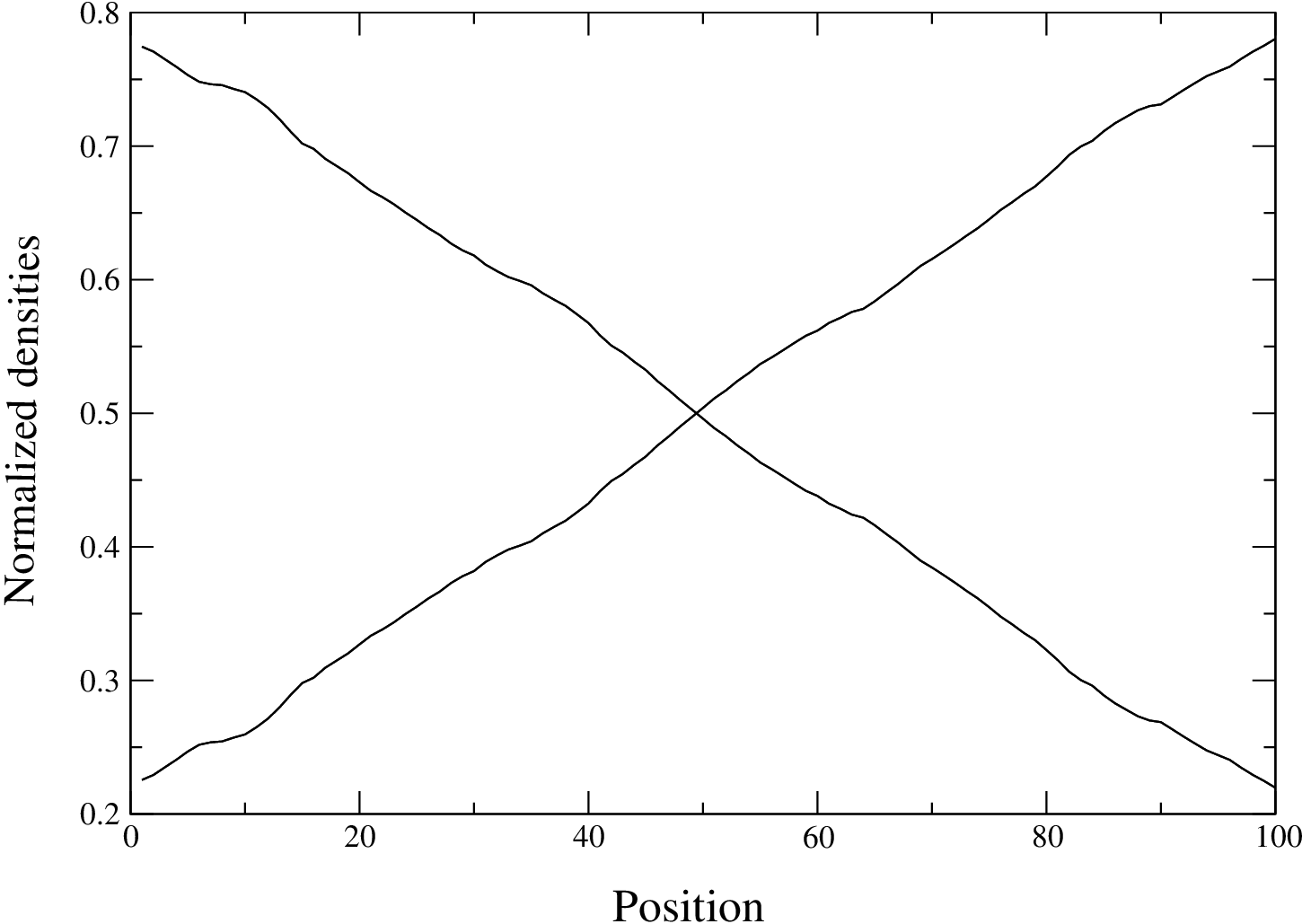}
\vskip 0.1in
\caption{Plot of the mass profile along the system for the particles
with mass $ m_+=1 $ when the other particles have a mass of $ m_-=0.
95 $. The length of the system is $ L=100 $. The temperature ratio
considered is $ Th/Tc $= 1.30. The curve with positive slope
correspond to $ m_+=1 $.
\label{fig9}}
\end{center}

\end{figure}

The fact that we have found the Fourier Law for the broad-interface
case is not surprising, in view of the similar result of Dhar \cite{Dhar}
 mentioned in the introduction. Note, however, that 
 unlike the present situation, in Ref. \cite{Dhar} 
 particles with different masses are never
 allowed to cross, so the Soret effect is avoided completely.
 
It can also be seen that for a fixed small thermal gradient, the
temperature profile depends linearly on the difference of masses. This
fact suggests the possibility of expanding the Boltzmann equation in
powers of this parameter and making an estimation of the conductivity
for the case in which the masses of the two species are very similar \cite{Fut} .

\section{CONCLUSIONS}

In this paper we have numerically solved a model with two types of
particles with different masses. We have shown that a thermal gradient
is enough to separate completely the two species, at least if the
system is large enough to accommodate the interface. 
The phase separation is due to the collision asymmetry (\ref{asy})
provoked by the thermal gradient applied. For large enough systems,
 the heavy particles are closer to the
hottest wall while the light particles are closer to the coldest
one. Between these two regions there is an interface where the two types
of particles live together.
 In the regions occupied by
only one type of particles the kinetic energy gradient is zero because the system
is a noninteracting gas, so that all the energy density drop is localized in the
interface (a finite region of the space).  
We have also studied the behavior of the interface for different
values of the characteristic parameters of the system, showing that
its instantaneous width  is independent of the system size.
By means of a numerical solution of the Boltzmann equation for this
systems, we have checked that the total phase separation is also
obtained within this approximation (equations
(\ref{boltz1}),(\ref{boltz2})).
We have also investigated the limit of `broad interface', in which the
interface is broader than the system itself, and shown that 
Fourier's Law is recovered in that case. 

To sum up, we have introduced a very simple one dimensional model 
which has an  extreme Soret effect for an infinite
system, while in a  finite system  reproduces not only
Fourier's Law but also a linear distribution of densities along the system.

\begin{center}
{\bf ACKNOWLEDGMENTS}
\end{center}

 We wish to thank B. Derrida, J. Lebowitz, M. Martin, R. Livi,
M. Rubi, S. Ruffo,  and J-E Wesfreid  for useful suggestions and discussions.
We also thank I. Pagonabarraga for a careful reading of
the manuscript.

\end{document}